\documentclass[namedreferences]{SolarPhysics}

\usepackage{graphicx}        
\usepackage{amssymb}        
\usepackage{url}             
\newcommand{\beq}{\begin{equation}}
\newcommand{\eeq}{\end{equation}}
\newcommand{\eq}[1]{(\ref{#1})}
\newcommand{\fb}{}

\newcommand{\adv}{    {\it Adv. Space Res.}}

\newcommand{\aap}{    {\it Astron. Astrophys.}}

\newcommand{\apj}{    {\it Astrophys. J.}}
\newcommand{\apss}{   {\it Astrophys. Space Sci.}}

\newcommand{\solphys}{{\it Solar Phys.}}

\newcommand{\ssr}{    {\it Space Sci. Rev.}}

\def\copyrightline{}

\begin{document}
\begin{article}
\begin{opening}
\newcommand\mytit{Solar cycle 25 prediction using length-to-amplitude relations}
\title{\mytit}
\author{Vladimir~G.~\surname{Ivanov}$^{1,2}$}
\runningauthor{Ivanov}
\runningtitle{\mytit}
\institute{$^{1}$Central Astronomical Observatory at Pulkovo, Saint-Petersburg, Russia\\
$^2$email: \url{vg.ivanov@gaoran.ru}}

\begin{abstract}
We propose a simple method for prediction of the 11-year solar cycle
maximum that is based on two relations.
One of them is the well known Waldmeier rule that binds the
amplitude of a cycle and the length of its ascending phase.
The second rule relates the length of a given cycle from minimum to minimum
and the amplitude of the next one. Using corresponding
linear regressions we obtain for the amplitide of cycle 25
in the scale of 13-month smoothed monthly total revised sunspot number 
${\rm SN}_{\rm max}(25) = 181\pm46$ and for the moment of the maximum $t_{\rm max}(25) = 2024.2\pm1.0$.
Therefore, according to the prediction, cycle 25 will be higher than
the previous one (${\rm SN}_{\rm max}(24) = 116$) with probability 0.92.
\end{abstract}
\end{opening}

\section{Introduction}

In 2019 experts of Solar Cycle 25 Prediction Panel made an assumption%
\footnote{\url{https://www.weather.gov/news/190504-sun-activity-in-solar-cycle}}
that the maximum of the 11-year solar cycle 25 (SC25) will lie in the sunspot
number range between 95 and 130, that is similar to SC24 with the
maximum $SN_{\rm max}(24)=116$.
The opinion that SC25 will be equal or lower than SC24 is shared by many authors 
(\opencite{du22}; \opencite{burud21}; \opencite{nandy21}; \opencite{wu21}; \opencite{courtillot21} etc).

If the odd SC25 is lower than the even SC24, it will violate
the Gnevyshev-Ohl correlation rule \cite{gnevyshev48}. Many regard
this as a sign that the Sun is entering a Dalton-like global
minimum of activity.

However, there is an alternative point of view that the forecast of the experts has been underestimated
and the forecoming cycle will be higher
(\opencite{macintosh20}; \opencite{koutchmy21}; \opencite{prasad22}; \opencite{lu22} etc).
None of the two positions have prevailed so far. 

In the following we propose a simple method of solar maximums
prediction based on linear relations between amplitudes of cycles and
lengths of {their} phases. This method provides one more argument in favor
of viewpoint that SC25 will be higher {than} SC24 and comparable with
SC23.

\section{Data and notation}

Hereafter we will use for analysys and prediction the 13-month smoothed monthly averages
of the recalibrated sunspot number SN for 1749--2021 \cite{clette14}%
\footnote{The dataset is available at \url{https://wwwbis.sidc.be/silso/DATA/SN_ms_tot_V2.0.txt}}.

We introduce the following notation: ${\rm SN}_{\rm max}(i)$ is the sunspot index in the maximum of the $i$th cycle
(i.e. its amplitude),
$t_{\rm min}(i)$ and $t_{\rm max}(i)$ are moments of its minimum and maximum,
$T_{\rm mm}(i) = t_{\rm min}(i+1) - t_{\rm min}(i)$ is its length (from minimum to minimum),
$T_{\rm mM}(i) = t_{\rm max}(i) - t_{\rm min}(i)$ is the length of the
ascending branch of the cycle (from minimum to maximum).

\section{Links between parameters of 11-year cycles}

\subsection{The Waldmeiers rule}

\begin{figure}
\begin{center}
\fb{\includegraphics[width=0.6\textwidth]{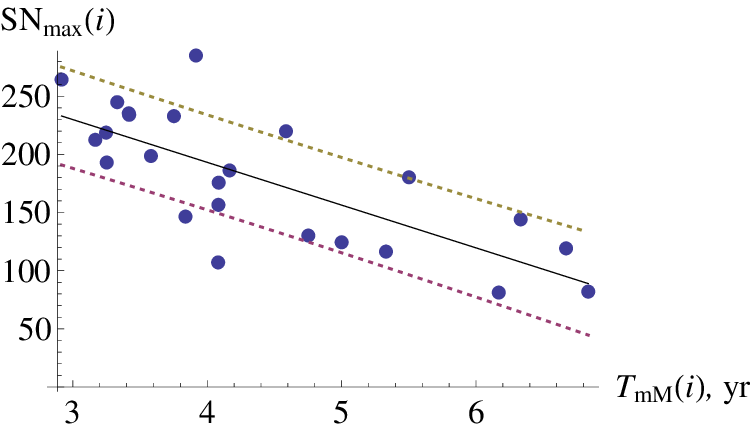}}
\end{center}
\caption{The relationship between the length of the ascending phase $T_{\rm mM}(i)$ and the amplitude of the cycle
${\rm SN}_{\rm max}(i)$ (the Waldmeier rule).
The solid line is the linear regression \eq{eqwr}.
The dashed lines mark the standard error intervals for predictions.}
\label{figwr}
\end{figure}

The well-known empirical Waldmeier rule (WR) states that the length
of the ascending branch of the cycle $T_{\rm mM}$ anticorrelates with its amplitude
${\rm SN}_{\rm max}$ \cite{waldmeier35} (Fig.~\ref{figwr})
{with the Pearson correlation coefficient $r(T_{\rm mM},{\rm SN}_{\rm max})=-0.75$}.
One can find the linear regression equation
\begin{eqnarray}
 {\rm SN}_{\rm max}(i) &=& a_1 + b_1\,T_{\rm mM}(i)\,,\nonumber\\
 a_1 &=&340.2 \pm 31.3\,,\nonumber\\
 b_1 &=& (-36.76 \pm 6.90)\;{\rm yr}^{-1}\,,
 \label{eqwr}
\end{eqnarray}
{and the standard deviation of the regression residuals $\sigma^{\rm(res)}_1 = 39.0$.}
{The correlation coefficient between the parameters of the regression,
which one may need to evaluate errors of predictions, is}
\[
 r(a_1,b_1) =-0.976\,.
\]

\subsection{The length-to-next-amplitude rule (LNAR)}

\begin{figure}
\begin{center}
\fb{\includegraphics[width=0.6\textwidth]{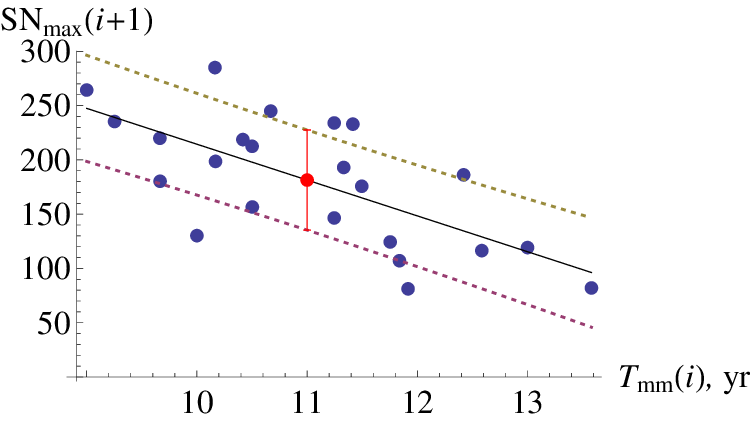}}
\end{center}
\caption{The relationship between the length of the cycle $T_{\rm mm}(i)$ and the amplitude of the next one
${\rm SN}_{\rm max}(i+1)$ (the LNA rule).
The solid line is the linear regression \eq{eqlnar}.
The red circle is the prediction for SC25.
The dashed lines mark the standard error intervals for predictions.}
\label{figlnar}
\end{figure}

Another empirical rule, similar to WR but less known (and still nameless, to the best of our knowledge), 
states that the length of a given cycle $T_{\rm mm}(i)$ anticorrelates with the amplitude
of the next one ${\rm SN}_{\rm max}(i+1)$ \cite{cher54, hathaway94, solanki02, hathaway15, macintosh20, ivanov21}
(Fig.~\ref{figlnar}). The corresponding regression is 

\begin{eqnarray}
 {\rm SN}_{\rm max}(i+1) &=& a_2 + b_2\,T_{\rm mm}(i)\,,\nonumber\\
 a_2 &=& 544.5 \pm 86.6\,,\nonumber\\
 b_2 &=& (-33.01 \pm 7.81)\;{\rm yr}^{-1}\,,\nonumber\\
 r(T_{\rm mm}(i),{\rm SN}_{\rm max}(i+1)) &=&-0.68\,,\nonumber\\
 \sigma^{\rm(res)}_2 &=& 44.1\,,\nonumber\\
 r(a_2,b_2) &=&-0.994\,.
 \label{eqlnar}
\end{eqnarray}

\subsection{The length-to-ascending-length rule (LALR)}

\begin{figure}
\begin{center}
\fb{\includegraphics[width=0.6\textwidth]{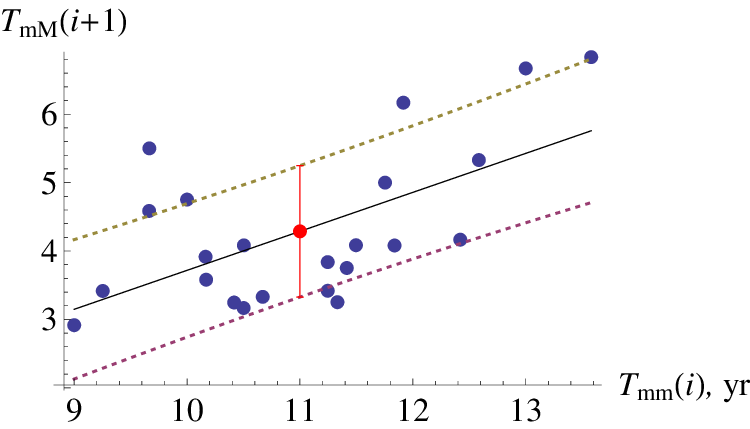}}
\end{center}
\caption{The relationship between the length of the cycle $T_{\rm mm}(i)$ and the length of the next ascending phase 
$T_{\rm mM}(i+1)$ (the LAL rule).
The red circle is the prediction for SC25.
The solid line is the linear regression \eq{eqlalr}.
The dashed lineses mark the standard error intervals for predictions.}
\label{figlalr}
\end{figure}

A direct consequence of the WR and LNAR is the third rule
that binds the length of a given cycle $T_{\rm mm}(i)$ with the length of the 
ascending phase of the next one $T_{\rm mM}(i+1)$ (Fig.~\ref{figlalr}).
These parameters correlate, and the regression is

\begin{eqnarray}
 T_{\rm mM}(i+1) &=& a_3 + b_3\,T_{\rm mm}(i) \,,\nonumber\\
 a_3 &=& (-1.98 \pm 1.81)\;{\rm yr}\,,\nonumber\\
 b_3 &=& 0.570 \pm 0.163\,,\nonumber\\
 r(T_{\rm mm}(i),T_{\rm mM}(i+1)) &=&+0.61\,,\nonumber\\
 \sigma^{\rm(res)}_3 &=& 0.92 \,,\nonumber\\
 r(a_3,b_3) &=& -0.994\,. 
 \label{eqlalr}
\end{eqnarray}

\section{Prediction of SC25 maximum}

{Linear regressions \eq{eqlnar} and \eq{eqlalr} based on} the LNA and LAL rules allow to estimate the moment and
magnitude of the maximum of the next cycle, {as well as prediction confidence intervals of the estimations (see, e.g., \opencite{sachs72})},
provided we know the length of the current one.
Since the minimum between SC24 and SC25 in the
13-month smoothed index occurs in December 2019, the length of the
SC24 is 11.0~yr, {we obtain predictions}
\[
{\rm SN}_{\rm max}^{(p)}(25)=181\pm46\,,
\]
\[
{\rm T}_{\rm mM}^{(p)}(25)=(4.23\pm0.96)\;{\rm yr}
\]
(see Figs.~\ref{figlnar} and~\ref{figlalr}), and, therefore,
\[
t_{\rm max}^{(p)}(25)=(2024.24\pm0.96)\;{\rm yr}\,,
\]
{where the errors correspond to the $1\sigma$ (68\%) confidence level.
Note that for rough estimate of the confidence intervals one can use 
the standard deviations of the regression residuals $\sigma^{\rm(res)}_2$ and $\sigma^{\rm(res)}_3$,
arriving at similar results.}

The errors of the predictions are large enough.
Nevertheless, {The probability for SC25 to be
higher than SC24 (${\rm SN}_{\rm max}(24) = 116$) is}
\[
\frac12\left[1+{\rm erf} \left( \frac{{\rm SN}_{\rm max}^{(p)}(25)-{\rm SN}_{\rm max}(24)}{\sqrt2\,\sigma_{{\rm SN}_{\rm max}}^{(p)}(25)} \right) \right]
= 0.92\dots
\,,
\]
{where ${\rm erf}(z)$ is the Gauss error function.}

\begin{figure}
\begin{center}
\fb{\includegraphics[width=0.6\textwidth]{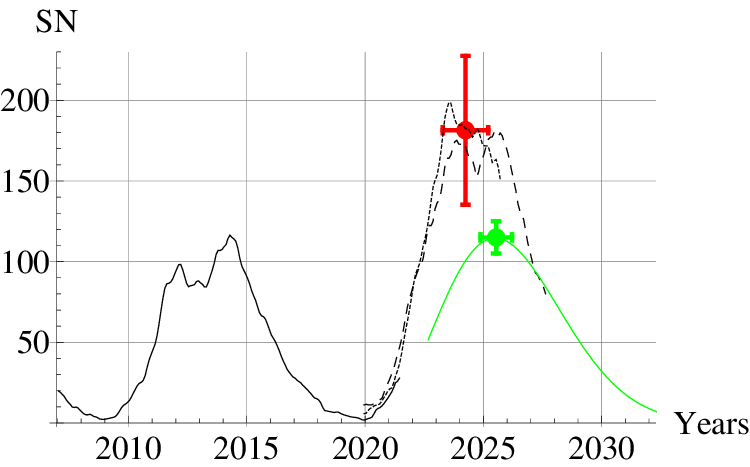}}
\end{center}
\caption{
The observed SN index for SC24 and SC25 (the solid black line),
SC17 (short dashes) and SC23 (long dashes) shifted to match the minimums
to that of SC25, our prediction of SC25 maximum (the red circle with errorbars),
the prediction of SC25 (the green line) and its maximum (green point with errorbars) by the Solar Cycle Prediction Panel
(the dataset available at \protect\url{https://www.swpc.noaa.gov/products/predicted-sunspot-number-and-radio-flux}).}
\label{figscxx}
\end{figure}

Probably, it will be similar to SC23 (${\rm SN}_{\rm max}(23)=176$) or SC17\linebreak
(${\rm SN}_{\rm max}(17)=189$) (Fig.~\ref{figscxx}). It means that
the odd SC25 will be higher than the even SC24, and the
Gnevyshev-Ohl rule will keep valid.

Earlier \cite{ivanov21}, using the same method, we have obtained
for SC25 weaker estimates: the amplitude $136\pm36$ and the moment of
maximum $2025.7\pm0.7$ (for the Gaussian smoothing of SN with
$\sigma=8$~months).
This underestimate can be explained by the fact
that in January 2021, when the mentioned article was being prepared, it
was difficult to determine the moment of SC25 minimum in the
smoothed index accurately, and we assumed it to occur too late,
in October 2020, when the last local minimum of the index before
its fast growth took place.

\section{Control of stability}

\begin{figure}
\begin{center}
\fb{\includegraphics[width=0.6\textwidth]{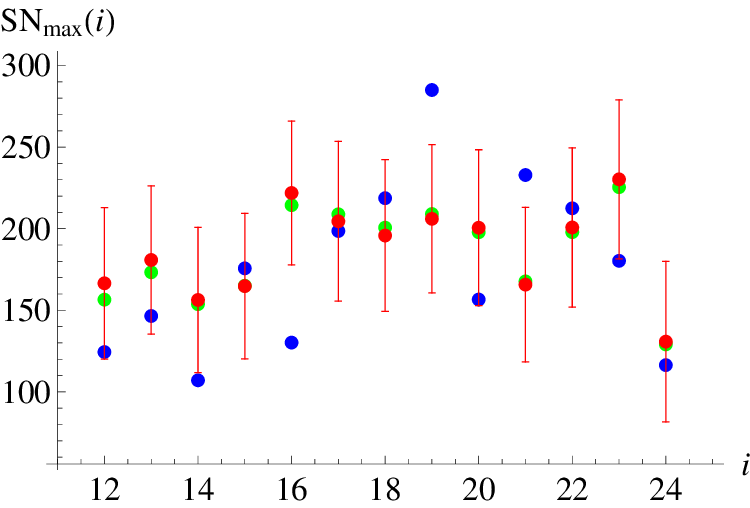}}
\fb{\includegraphics[width=0.6\textwidth]{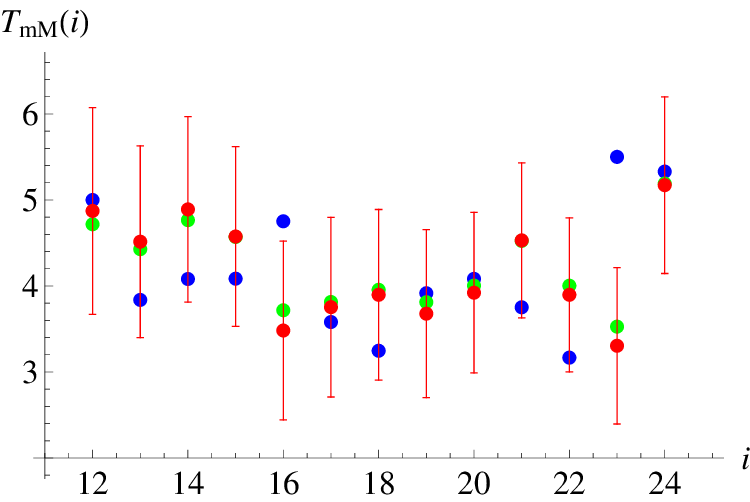}}
\end{center}
\caption{The control predictions for ${\rm SN}_{\rm max}$ (top) and $T_{\rm mM}$ (bottom)
on truncated series. The red circles with errorbars are predictions of the $i$th SC.
The green circles (which partly overlap with red) are predictions made by the full length series.
The blue circles are observed values.}
\label{figcamp}
\end{figure}

The parameters of the mechanism that drives the solar activity can
vary over time, affecting the rules under discussion and
efficiency of the described method of prediction. To control its
stability we will make the following procedure. Let's take the
subseries of parameters from 1st to $(i-1)$th cycle, construct the
regressions for the LNAR and LALR using the truncated series and
obtain the predictions for the moment and amplitude of the $i$th
cycle. The results of the predictions are plotted in
Figs.~\ref{figcamp}, where the observed values and
predictions made on the base of full series are also plotted. All
predicted values are in $1.5\sigma$ bands relative to the observed
values. The number of predictions in the $1\sigma$ band is 8 out of 13
(62\%) for amplitudes and 11 out of 13 (85\%) for lengths, which is
close to the that probability for the normal
distribution of errors (68\%).
In addition, one can see that predictions made on the
base of truncated and full series are very close (they differences do
not exeed $0.25\sigma$). These facts indicate stability of the
relations used for predictions at least in the last century.

\section{Dependence on smoothing}

{
Above we made a prediction for the 13-month running average of the SN series.
To study dependence of the results on the smoothing method we repeated the procedure for:
\begin{itemize}
\item the running average with the smoothing window width $w$ and weights
$$\frac{1}{2(w-1)}, \frac{1}{w-1}, \dots, \frac{1}{w-1}, \frac{1}{2(w-1)}$$
($w=13$ corresponds to the standard 13-month average);
\item the smoothing by the Gaussian kernel with window width $w$ and $\sigma=w/2$.
\end{itemize}
}

\begin{figure}
\begin{center}
\fb{\includegraphics[width=0.6\textwidth]{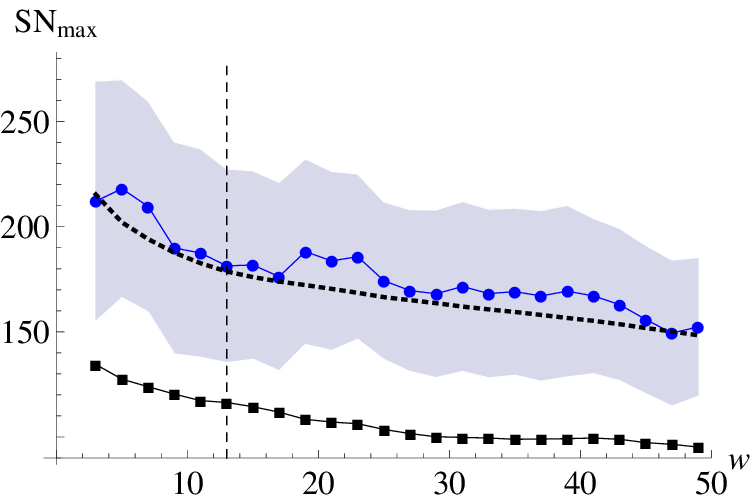}}
\fb{\includegraphics[width=0.6\textwidth]{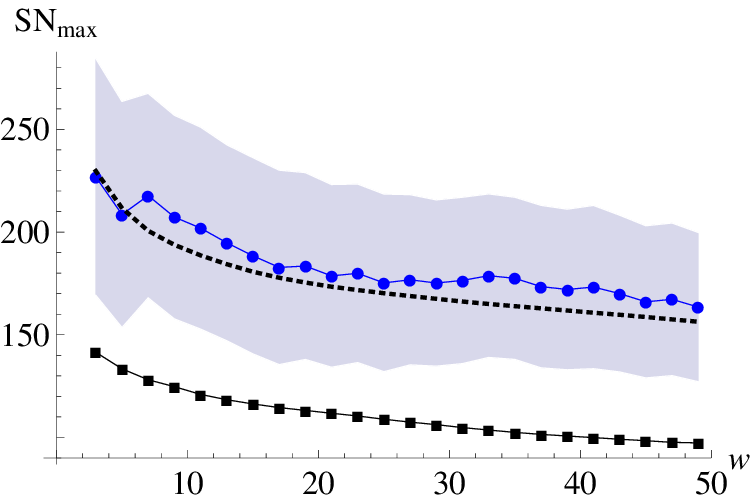}}
\end{center}
\caption{{The predictions of SC25 maximum vs. the width $w$ of the smoothing window (blue circles).
The grey filling corresponds to 1-sigma confidence bands of the prediction.
The maximums of SC24 for the same smoothing are shown by squares and the averaged maximums over cycles 1--24, by the dotted curve.
The upper panel corresponds to the case of running average and the bottom one, to the Gaussian smoothing.
The dashed vertical line marks the standard 13-month smoothing.}}
\label{fighigher}
\end{figure}

\begin{figure}
\begin{center}
\fb{\includegraphics[width=0.6\textwidth]{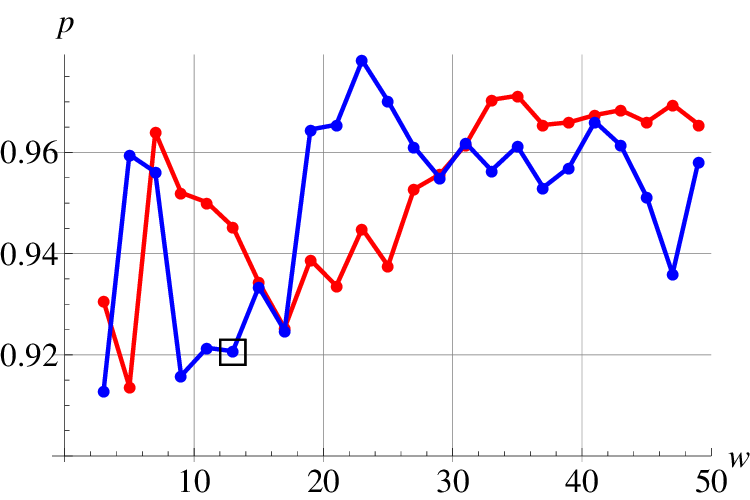}}
\end{center}
\caption{{The probability $p$ that SC25 will be higher than SC24 vs. the width of the smoothing window $w$
for the running average (blue) and Gaussian smoothing (red). The point in the box corresponds to the standard 13-month smoothing.}}
\label{figprob}
\end{figure}

{
In both cases $w$ is odd and varies from 3 to 49 months.
The results are presented in Figures~\ref{fighigher} and~\ref{figprob}.
We can see that, despite the amplitudes of cycles decrease with growth of
the smoothing scale, ratios of their heights keep approximately the same.
In particular, the predicted ${\rm SN}_{\rm max}(25)$ for any smoothing is
close to the amplitude averaged over cycles 1--24
and higher than ${\rm SN}_{\rm max}(24)$ with probability $p>0.91$.
}

\section{Conclusions}

The proposed method of prediction, despite its simplicity and
rather large errors, provides an accuracy of about
15--30\% for the moment of maximum and 25--50\% in its magnitude,
which is enough to distinguish between forthcoming cycles of low, medium and
high power. In particular, the method predicts SC25 of medium
magnitude ${\rm SN}_{\rm max}(25) = 181\pm46$, which will be
higher than the previous one (${\rm SN}_{\rm max}(24) = 116$) with
probability 0.92, and with the same probability
the Gnevyshev-Ohl correlation rule will be valid for SC25.

{
The relation between the length of the current cycle and the maximum of the next one
that we exploit here was also used for prediction of SC25 by \inlinecite{macintosh20}.
Using more sophysticated approach based on the Hilbert transformation of the SN series, they
defined a length of the 11-year cycle as the time between ``termination events''
and obtained a preliminary estimate for the maximum ${\rm SN}_{\rm max}(25) = 233^{+21}_{-29}$. However, the validity of such
approach is still questinable and its results may depend on details of the procedure
(see a discussion in \inlinecite{boot21} and \inlinecite{leamon21}).
In particalar, \citeauthor{macintosh20} have apparently underestimated the length of SC24 and, therefore,
overestimated the power of SC25 (in \inlinecite{leamon21} 
the same method arrived at a lower prediction ${\rm SN}_{\rm max}(25)=195\pm17$.
}

{
In our opinion, the advantage of the simplest approach, with the length of the 11-year cycle
defined as the time between two seccessive minimus is that the only detail that could effect
the results of prediction is a type and scale of smoothing. However, as we have shown above, it is not the case.
}

The prediction ${\rm SN}_{\rm max}(25) = 181\pm46$ is obtained for 
the 13-month smoothed series. Since the series ends six months earlier than the non-smoothed one,
it was possible to made it seven month after the minimum occurs, that is
in July 2020. {However, as Figures~\ref{fighigher} and~\ref{figprob} show, the qualitive conclusion about
moderate amplitude of SC25 that is higher than SC24 keeps valid for smaller smoothing scales,
including $w=3$~months, so it was possible to make that conclusion as early as in the beginning of 2020.}

\section{Data Availability}

The datasets analysed during the current study are openly available at resources referenced above.
The datasets generated during the study are available from the author on reasonable request.

\end{article}


\begin{thebibliography}{00}

\bibitem[\protect\citeauthoryear{{Booth}}{2021}]{boot21}
Booth R.J.: 2021, Limitations in the Hilbert Transform Approach to Locating Solar Cycle Terminators.
\textit{\solphys} \textbf{296}, 108.

\bibitem[\protect\citeauthoryear{{Burud et al.}}{2021}]{burud21}
Burud, D.S., Jain, R,, Awasthi, A.K., Chaudhari, S., Tripathy, S.C., Gopalswamy, N., Chamadia, P., Kaushik, S,C., Vhatkar, R.: 2021,
Spotless days and geomagnetic index as the predictors of solar cycle 25.
\textit{Research in Astronomy and Astrophysics} \textbf{21}, 215.

\bibitem[\protect\citeauthoryear{{Chernosky}}{1954}]{cher54}
Chernosky, E.J.: 1954, A relationship between the length and activity of sunspot cycles,
\textit{Publ. Astron. Soc. Pac.} \textbf{66}(392), 241.

\bibitem[\protect\citeauthoryear{{Clette et al.}}{2014}]{clette14}
Clette, F., Svalgaard, L., Vaquero, J.M., Cliver, E.W.: 2014,
Revisiting the sunspot number.
\textit{\ssr} \textbf{186}, 35.

\bibitem[\protect\citeauthoryear{{Courtillot et al.}}{2021}]{courtillot21}
Courtillot, V., Lopes, F., Le Mou\"el, J.L.: 2021,
On the Prediction of Solar Cycles.
\textit{\solphys} \textbf{296}, 21.

\bibitem[\protect\citeauthoryear{{Du et al.}}{2022}]{du22}
Du, Z.L.: 2022,
The solar cycle: a modified Gaussian function for fitting the shape of the solar cycle and predicting cycle 25.
\textit{\apss} \textbf{367}, 20.

\bibitem[\protect\citeauthoryear{{Gnevyshev and Ohl}}{1948}]{gnevyshev48}
Gnevyshev, M.N., Ohl, A.I.: 1948,
On 22-year cycle of the solar activity.
\textit{Astron. Zh.} \textbf{25}, 18.

\bibitem[\protect\citeauthoryear{{Hathaway}}{2015}]{hathaway15}
Hathaway, D.H.: 2015,
The Solar Cycle.
\textit{Living Rev. Sol. Phys.} \textbf{12}, 4.

\bibitem[\protect\citeauthoryear{{Hathaway et al.}}{1994}]{hathaway94}
Hathaway, D.H., Wilson, R.M., Reichmann, E.J.: 1994,
The shape of the sunspot cycle.
\textit{\solphys} \textbf{151}, 177.

\bibitem[\protect\citeauthoryear{{Ivanov}}{2021}]{ivanov21}
Ivanov, V.: 2021,
Two Links between Parameters of 11-year Cycle of Solar Activity.
\textit{Geomagnetism and Aeronomy} \textbf{61}, 1029.

\bibitem[\protect\citeauthoryear{{Koutchmy et al.}}{2021}]{koutchmy21}
Koutchmy, S., Tavabi, E., No\"ens, J.-C., Wurmser, O., Filippov, B.: 2021,
Predicting the height of the solar cycle 25 through polar regions activity.
In: Siebert A. et al. (eds.) \textit{Proceedings of the Annual meeting of the French Society of
Astronomy and Astrophysics}, 238.

\bibitem[\protect\citeauthoryear{{Leamon et al.}}{2021}]{leamon21}
Leamon, R.J., McIntosh, S.W., Chapman, S.C., Watkins, N.W.: 2021,
Response to ``Limitations in the Hilbert Transform Approach to Locating Solar Cycle Terminators'' by R. Booth 
\textit{\solphys} \textbf{296}, 151.

\bibitem[\protect\citeauthoryear{{Lu et al.}}{2022}]{lu22}
Lu, J.Y., Xiong, Y.T., Zhao, K., Wang, V, Li, J.Y., Peng, G.S., Sun, M.: 2022, 
A Novel Bimodal Forecasting Model for Solar Cycle 25.
\textit{\apj} \textbf{924}, 59.

\bibitem[\protect\citeauthoryear{{McIntosh et al.}}{2020}]{macintosh20}
McIntosh, S.W., Chapman, S.C., Leamon, R.J., Egeland, R., Watkins, N.W.: 2020,
Overlapping magnetic activity cycles and the sunspot number: forecasting sunspot cycle 25 amplitude.
\textit{\solphys} \textbf{295}, 163.

\bibitem[\protect\citeauthoryear{{Nandy}}{2021}]{nandy21}
Nandy, D.: 2021,
Progress in Solar Cycle Predictions: Sunspot Cycles 24-25 in Perspective.
\textit{\solphys} \textbf{296}, 54.

\bibitem[\protect\citeauthoryear{{Sachs}}{1972}]{sachs72}
Sachs L.: 1972, Statistische Auswertungsmethoden, Springer-Verlag, Berlin -- Heidelberg -- New York.

\bibitem[\protect\citeauthoryear{{Solanki et al.}}{2002}]{solanki02}
Solanki, S.K., Krivova, N.A., Sch\"ussler, M., Fligge, M.: 2002,
Search for a relationship between solar cycle amplitude and length.
\textit{\aap} \textbf{396}, 1029.

\bibitem[\protect\citeauthoryear{{Prasad et al.}}{2022}]{prasad22}
Prasad, A., Roy, S., Sarkar, A., Chandra Panja, S., Narayan Patra, S.: 2022,
Prediction of solar cycle 25 using deep learning based long short-term memory forecasting technique.
\textit{\adv} \textbf{69}, 798.

\bibitem[\protect\citeauthoryear{{Waldmeier}}{1935}]{waldmeier35}
Waldmeier, M.: 1935,
Neue Eigenschaften der Sonnenfleckenkurve.
\textit{Astron. Mitt. Eidgen\"ossischen Sternwarte Z\"urich}
\textbf{14}, 105.

\bibitem[\protect\citeauthoryear{{Wu and Qin}}{2021}]{wu21}
Wu, S.-S., Qin, G.: 2021,
Predicting Sunspot Numbers for Solar Cycles 25 and 26.
E-print arXiv:2102.06001.

\end{thebibliography}
\end{document}